\documentclass[twocolumn,pre]{revtex4}
\usepackage{graphicx}
\usepackage[dvips]{color}
\begin{document}

\title{Nondiffusive heat transfer in muscle tissue. Preliminary results}
\author{E. V. Davydov}
\author{I. A. Lubashevsky}
\email[address for contacts: ]{ialub@fpl.gpi.ru}
\author{V. A. Milyaev}
\author{R. F. Musin}
\affiliation{General Physics Institute, Russian Academy of Sciences, Vavilov
str. 38, Moscow, 117942, Russia}
\date{\today}

\begin{abstract}
We present preliminary experimental data that enable us to suggest that heat
transfer in cellular tissue under local strong heating is a more complex
phenomenon than a simple heat diffusion. Namely, we demonstrate that under
local strong heating of a muscle tissue heat transfer in it exhibits
substantial anisotropy unexplained in the context of the standard diffusion
model. The observed temperature dynamics is also characterized by nonlinear
behavior as well as by a certain repeat reversibility. The latter means that
the time variations in the temperature of a cellular tissue undergoing repeated
acts of heating go in the same way at least approximately. We explain the
observed anomalous properties of heat transfer by suggesting the flow of the
interstitial liquid to appear due to nonuniform heating which, in turn, affects
the heat transfer. A possible mechanism responsible for this effect is
discussed.
\end{abstract}
\maketitle

\section{Introduction\protect\label{sec:intro}}

During the last two decades the description of heat transfer in living tissue
(bioheat transfer) and mathematical modeling of temperature distribution in it
under local strong heating have attracted attention due to several reasons.
First, this problem is an important part of planning and optimizing
hyperthermia treatment and thermotherapy of tumors where the precise control
over the tissue temperature is essential (see, e.g., \citep{I1,I2,I3}).
Second, what is interesting from the standpoint of fundamental physics, living
tissue is a complex medium in which highly branching blood vessels form a
fractal network of fast heat transport \cite{I4}, endowing heat propagation in
it with anomalous properties \cite{BLG}. Third, living tissue is an active
medium, in particular, it responds to strong heating, which causes the blood
perfusion rate to grow locally by tenfold \cite{Song84}. In this case one
meets also the basic problem of how a natural hierarchical system comprising a
huge amount of elements can respond properly to changes in its state when none
of the elements possesses the total information required to govern the system
as a whole \cite{LGenv95,gaf1}.

By now a number of significant steps towards developing the bioheat transfer
theory have been made. In particular, Chen \& Holmes~\cite{CH80} actually
stated the microscopic description of the problem. Baish \textit{et
al.}~\cite{Bi86}, Lagendijk \textit{et al.}~\cite{LD1,LD2}, Roemer \textit{et
al.}~\cite{RD1,RD2}, and Gafiychuk \& Lubashevsky \cite{BLG} studied the effect
of the vessel discreteness on the temperature field. Weinbaum \textit{et al.}
\cite{WJ85,WXZE97} and Lubashevsky \& Gafiychuk \cite{BLG} took into account
sophisticated details of heat interaction between different vessels, namely,
the counter-current effect, Chato \cite{C80} and Weinbaum \& Jiji \cite{WJ85},
that renormalizes the kinetic coefficients of heat propagation in the tissue.
Lubashevsky \& Gafiychuk \cite{BLG,LGenv95,gaf1} proposed a hierarchical model
for the tissue response to local heating. There is also an approach by
Lagendijk \textit{et al.} (see, e.g., \cite{Lag} and references therein) that
dealing with individual vessels tackles the bioheat transfer problem
numerically.

All these models and approaches are based on the assumption that the cellular
tissue is a \textit{passive} medium and heat propagation in it obeys Fick's
law, i.e. the heat flux $\mathbf{Q}$ in the cellular tissue is proportional to
the temperature gradient $\nabla T$. Namely, $\mathbf{Q}=\kappa\nabla T$ and
the thermal conductivity $\kappa$ is isotropic or, at least, is slightly
anisotropic. Indeed, according to the experimental data summarized in
\cite{GMLM95}, for example, the value of the temperature diffusivity
$D=\kappa/(c\rho)$ ($c$ and $\rho$ are the specific heat and density of the
tissue) is approximately the same for different tissues, $D\cdot 10^3\approx
1.8$, 1.3, 1.4, 1.5,  1.2 cm${}^2$/s for human muscle, kidney, spleen, liver,
and blood, respectively. We note that models developed in food engineering
also regard the cellular tissue as a passive medium (see, e.g., \cite{FE}).
This assumption is justified by a number of self-consistent results obtained
in a variety of experimental works devoted to measuring thermal
characteristics of perfused and nonperfused biological tissues (see, e.g.,
\cite{GMLM95} and references therein, and, also, \cite{ExC,V1,V2,V3,GrE}).
However, measuring thermal characteristics of living tissue is a difficult task
because of its complex structure. So, interpretation of the obtained
experimental data and verification of the corresponding mathematical models
are related problems which are far from being completed (see, e.g.,
\cite{LX,Var}), concerning especially strong heating. Nevertheless, up to now
all the peculiarities of heat transfer in living tissue are assumed to be due
to the convective transport with blood.

During hyperthermia treatment the tissue temperature grows up to the upper
boundary of the tissue surviving, i.e. up to $44-46\,{}^{\circ}$C, moreover,
during a thermotherapy course based on the thermal tissue coagulation the
temperature can attain values about 100\,${}^{\circ}$C. So the question of
whether the cellular tissue undergoing so strong heating can be described in
the framework of the passive medium model is not evident. Therefore the
thermal properties of cellular tissue under local strong heating deserve
individual investigations to clarify if the physical background of the bioheat
transfer theory has to be modified. In particular, the fact that heat
propagation in cellular tissue can exhibit anomalous properties was noted in
\cite{HDE}.

In this paper we present preliminary experimental data that have enabled us to
doubt whether cellular tissue, at least some of its types, is in fact a
passive medium and heat propagation in it does obey Fick's law.

\section{Temperature dynamics in the tissue bulk under strong local
heating\protect\label{sec:bulk}}

\begin{figure}
\begin{center}
\includegraphics[width=75.8mm,height=62.1mm]{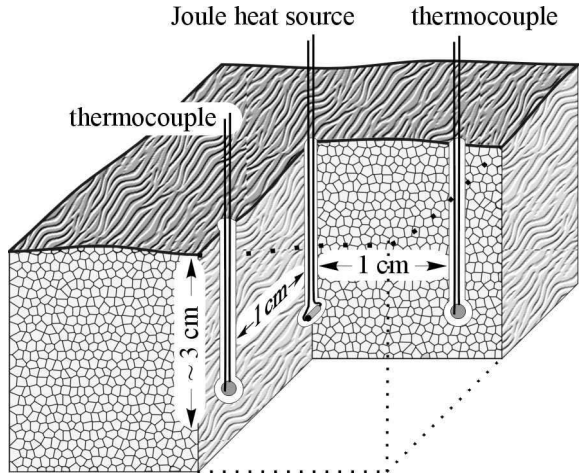}
\end{center}
\caption{Scheme of measuring the tissue temperature inside tissue bulk under
local heating caused by current running through the resistance.\label{F1}}
\end{figure}

First, we studied time variations of the temperature inside a fragment of cow
haunch muscle not undergone freeze after death and cut out so that muscle
fibers be parallel to its surface (Fig.~\ref{F1}). A small resistance $r$ of
size $1\times 2$\,mm was embedded into the tissue at a depth of 3\,cm through
this free surface oriented horizontally. Passing a current $I$ through it we
got a local heat source of controlled power. Thermocouples measuring the
temperature were located at a distance $\rho$ of 1\,cm from the resistance
along the fibers and in the transverse direction.

\begin{figure}
\begin{center}
\includegraphics[width=75.8mm,height=62.1mm]{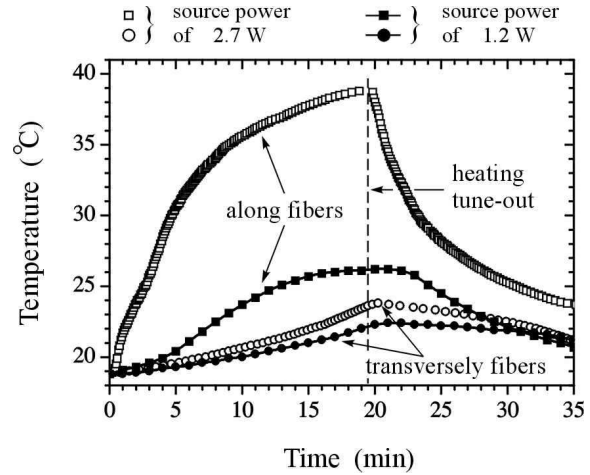}
\end{center}
\caption{Time variations of the tissue temperature at points shown in
Fig.~\protect\ref{F1} for two values of heating power. The data have been
obtained for tissue regions heated for the first time. The curves corresponding
to heating power of 1.2\,W are shifted down by 1.4\,${}^{\circ}$C to make the
comparison more easy.\label{F2}}
\end{figure}

We studied the heating dynamics for various values of the initial tissue
temperature from 0\,${}^{\circ}$C to 30\,${}^{\circ}$C and various specimens of
muscle, the results were practically identical. Fig.~\ref{F2} demonstrates
typical time variations of the tissue temperature obtained for two values of
the heat source power equal to 1.2\,W ($I=300$\,mA, $r=13$\,O) and 2.7\,W
($I=300$\,mA, $r=30$\,O). The time origin, $t = 0$, is the beginning of
heating and the vertical dashed line, $t\approx 20$\,min, represents the
heating tune-out. All the shown curves were obtained for tissue regions
undergone heating for the first time because under strong heating irreversible
transformations of the living tissue structure can occur. It should be noted
that the curves corresponding to heating power of 1.2\,W were obtained for the
tissue with the initial temperature $T_{\mathrm{init}}=20.2\,{}^{\circ}$C, so
in Fig.~\ref{F2} they are shifted down by 1.4\,${}^{\circ}$C to make the
comparison more easy.

As seen in Fig.~\ref{F2} the time variations of the tissue temperature along
the fibers and in the transverse direction exhibit substantially different
features. In fact, for the longitudinal direction the characteristic time
$\tau_l$ of the temperature variations can be estimated as $\tau_l\sim
6-8$\,min from the $e$-fold decrease after the heating tune-out. The same
estimate will be get if we apply to the point where the curve describing the
temperature increase under 1.2\,W heating transfers from the concave to convex
form. For the temperature increase under 2.7\,W heating the latter time seems
to be even shorter. The analogous curves describing the temperature variations
in the transverse direction are characterized by a substantially longer time
scale $\tau_t$. In particular, the temperature increase curves remain concave
up to the heating tune-out. So, if we related the time scales $\tau_{l,t}$ to
the temperature diffusivities $D_{l,t}$ along the fibers and transversely them
with the expression $\tau_{l,t}\sim \rho^2/D_{l,t}$ then we would have been to
set that $D_l/D_t\gtrsim 3$. From our point of view the latter estimate
contradicts the physical properties of heat transfer in cellular tissue
because it does not contain any regions where the heat conductivity takes so
different values that could give rise to this estimate.

Under strong heating the cellular tissue coagulates and changes its properties,
however, on the time scales under consideration it is essential only for
temperatures exceeding $50-60\,{}^{\circ}$C \cite{J}. Therefore, this effect
seems not to be substantial enough to play a key role in the heat propagation
through the cellular tissue in the given case. Indeed, if we again assume the
tissue temperature to be governed by the diffusion equation with, may be, an
anisotropic diffusivity, then the stationary temperature field $T$ caused by a
$\delta$-source will be scaled up as $(T-T_{\mathrm{init}})\rightarrow
(T-T_{\mathrm{init}})\lambda$ when all the spatial scales are reduced by the
factor $\lambda$, i.e. $\rho\rightarrow \rho/\lambda$. So, under 2.7\,W
heating the tissue temperature should attain values about 60\,${}^{\circ}$C at
points being at a distance of 0.5\,cm from the heat source along the fibers
and at a substantially shorter distance in the transverse direction. Thus the
region where the cellular tissue has to coagulate during the heating is of the
relative volume not exceeding 10\% with respect to the
1$\times$1$\times1$\,cm${}^3$ region actually tested in the experiment. Under
1.2\,W heating the region of the coagulated tissue is still less and even seems
not to form at all because of finite dimensions of the resistance.

Fig.~\ref{F2} also shows that heat propagation in cellular tissue under strong
heating must be nonlinear. At least, the change of the heating power from
1.2\,W to 2.7\,W, i.e. by 2.3~times, corresponds to the temperature increase
by 2.7~times at the points distant from the heat source along the fibers and
only by 1.4~times at the points in the transverse direction.

\begin{figure}
\begin{center}
\includegraphics[width=69mm,height=59.5mm]{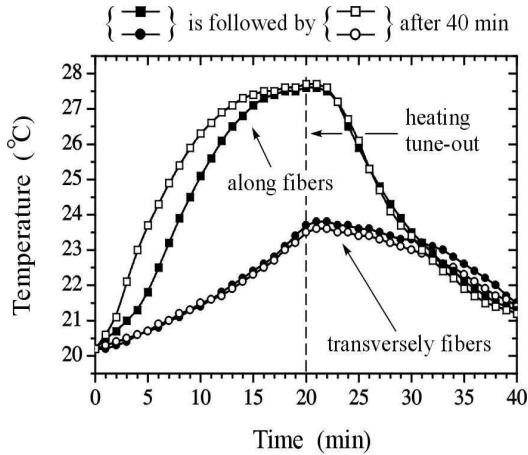}
\end{center}
\caption{Time variations of the tissue temperature at points shown in
Fig.~\protect\ref{F1} for the cellular tissue undergone two-fold heating of
power of 1.2\,W.\label{F3}}
\end{figure}

Fig.~\ref{F3} demonstrates our test of the repeat irreversibility of the
heating process showing if irreversible transformations of the tissue structure
caused by strong heating are essential in the case under consideration.
Namely, to answer the question of whether heat transfer in cellular tissue
exhibiting such anomalous properties can go in the same way under repeated
acts of heating we tested the temperature dynamics for the source power of
1.2\,W. The obtained curves present time variations of the tissue temperature
at the points shown in Fig.~\ref{F1} during two successive processes of
heating of the same tissue region that are separated in time by 40\,min. We see
not too significant change in the temperature variations only at the point
distant from the heat source along the fibers that occurs at the stage of
direct heating. During the repeated heating the temperature increase is
characterized solely by a faster growth rate. So the mechanism responsible for
these anomalous properties of heat transfer seems to be of reversible nature,
at least, at first approximation.

We assume that these heat transfer anomalies are due to motion of the
interstitial liquid induced by heating, which is demonstrated in the next
section.

\section{Temperature dynamics in a tissue layer under strong surface
heating\protect\label{sec:layer}}

\begin{figure}
\begin{center}
\includegraphics[width=69mm,height=59.5mm]{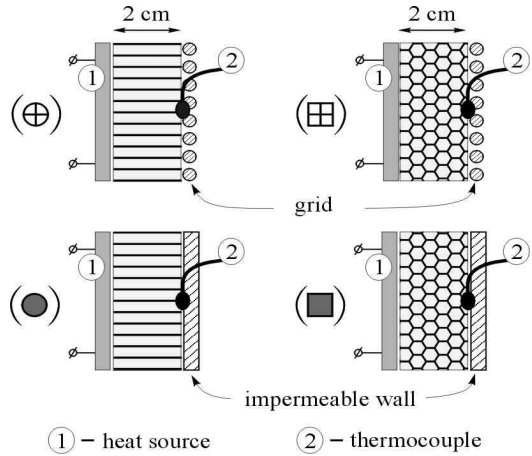}
\end{center}
\caption{Scheme of testing heat propagation through a tissue layer bounded by
wall permeable (upper figures) and impermeable (lower figures) to the
interstitial liquid.\label{F4}}
\end{figure}

Following actually a standard way of measuring the tissue heat conduction we
studied heat propagation through a tissue layer of thickness of 2\,cm
(Fig.~\ref{F4}). One of its surfaces was kept up at high temperatures about
100\,${}^{\circ}$C whereas the temperature growth at the opposite surface was
recorded by a thermocouple. The tissue layer was oriented vertically to avoid
the gravitation effect. We studied four arrangements distinguished by the fiber
orientation with respect to the walls bounding the layer and by the wall
permeability to the interstitial liquid. In one case we used the impermeable
walls to suppress the induced motion of the interstitial liquid, in the other
we applied a grid to make it as free as possible. The distance between the
walls was controlled by a certain small fixed force pressing the system so that
there were no air-gaps between the tissue and the walls.

It should be noted that in the given situation heat propagation through the
cellular tissue differs substantially from the case considered in the previous
section. In fact, when the thickness of the tissue layer is much less than the
other dimensions and its surface is heated uniformly, the temperature field
can practically vary solely along the direction perpendicular to the layer
surface. Under such conditions the interstitial liquid flow can appear only in
the same direction and inevitably has to be affected by the properties of the
opposite layer surface. For local heating of the cellular tissue at internal
points distant far from its surface the structure anisotropy should give rise
to the interstitial liquid flow of partly swirling form. In this case the
effect of the surface properties seems to be weakened.

\begin{figure}
\begin{center}
\includegraphics[width=72.7mm,height=59.7mm]{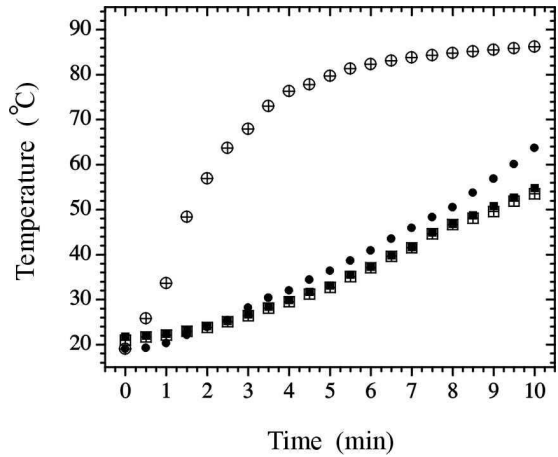}
\end{center}
\caption{Time variations of the temperature at the boundary of the tissue layer
opposite to the heated one (Fig.~\protect\ref{F4}). To make the relation
between the specific data and the corresponding tissue-wall configuration
clear the latter is labeled in Fig.~\protect\ref{F4} with the same
symbol.\label{F5}}
\end{figure}

Fig.~\ref{F5} exhibits the obtained results. When the tissue layer is bounded
by the impermeable walls and the interstitial liquid has no way to move the
evolution of tissue temperature is to be governed by the conventional heat
diffusion and to depend weakly on the fiber orientation. Exactly such a
behavior was observed. Then assuming the tissue temperature to obey the
diffusion equation we can directly estimate the corresponding temperature
diffusivity $D$. Namely, in the given case the tissue temperature at one
boundary is fixed and equal approximately to 100\,${}^{\circ}$C and at the
other we may impose the zero flux condition on the temperature field. The
initial temperature is 20\,${}^{\circ}$C and according to Fig.~\ref{F5} the
tissue temperature attains values about 50\,${}^{\circ}$C at latter boundary
in 10~min. Then fitting the solution of the diffusion equation subject to the
appropriate conditions to these data we get the estimate $D\sim 2\times
10^{-3}$\,cm${}^2$/s. The latter corresponds well to the available data for the
tissue temperature diffusivity within the accuracy of our measurements.

In the tissue layer bounded by the grid heat transfer exhibits another
behaviour. When the muscle fibers were parallel to the layer boundaries the
temperature growth turned out to be of the same form as for the impermeable
walls. By contrast, when the tissue fibers were perpendicular to the
boundaries the temperature growth deviated substantially from the previous
dependence (Fig.~\ref{F5}). We can explain this effect only admitting an
essential flux of the interstitial liquid to appear due to the nonuniform
heating of the cellular tissue. It is likely that the fiber arrangement gives
rise to long channels in the muscle tissue running along the fibers. So, when
there are no substantial obstacles to motion of the interstitial liquid
nonuniform heating of the tissue is able to induce the interstitial liquid
flux affecting remarkably heat transfer. To partly justify the given assumption
we note that for the latter fiber orientation the blood efflux through the grid
was visually strong in comparison with the other cases when the measured tissue
temperature attained the values about 33\,${}^{\circ}$C.

In order to make this phenomenon more evident the next section presents its
certain visualization.

\section{Mass propagation caused by nonuniform heating\protect\label{sec:visual}}

If nonuniform heating of cellular tissue causes the interstitial liquid to flow
in it affecting substantially heat propagation itself then the interstitial
liquid flux can be directly visualized through the induced propagation of a
certain substance dying the tissue. Indeed, diffusion coefficients of simple
molecules in cellular tissue (of order 10${}^{-5}$\,cm${}^2$/s or even less)
are much less than the temperature diffusivity, so mass propagation in
cellular tissue is to be essentially accelerated by this interstitial liquid
flux.

\begin{figure}
\begin{center}
\includegraphics[width=73mm,height=40.7mm]{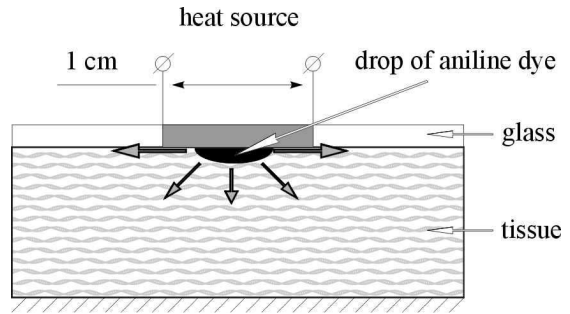}
\end{center}
\caption{Scheme of the experiment demonstrating the mass propagation induced by
nonuniform heating of the cellular tissue.\label{F6}}
\end{figure}

\begin{figure}
\begin{center}
\includegraphics[width=80mm,height=60mm]{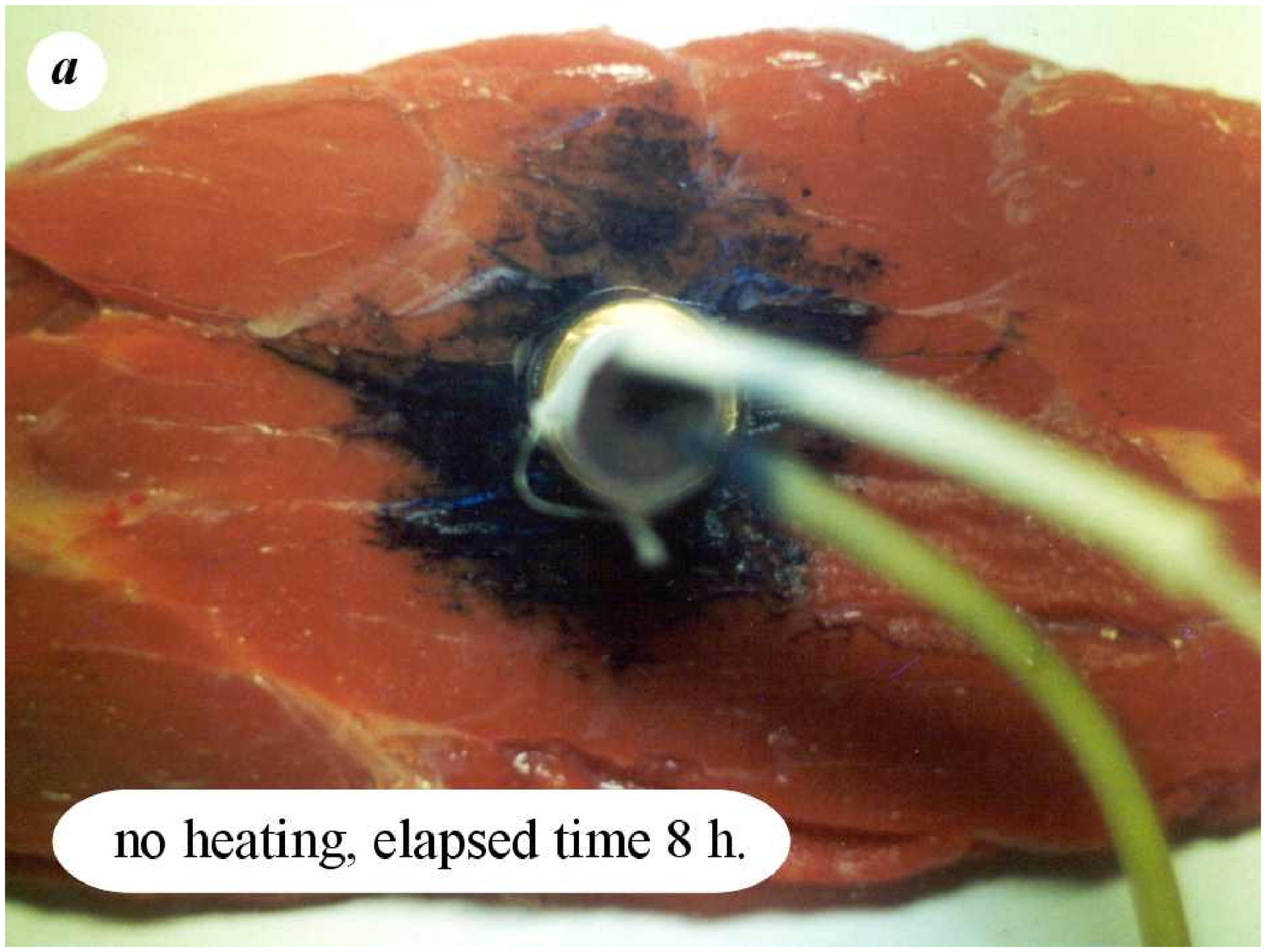}
\end{center}
\begin{center}
\includegraphics[width=80mm,height=60mm]{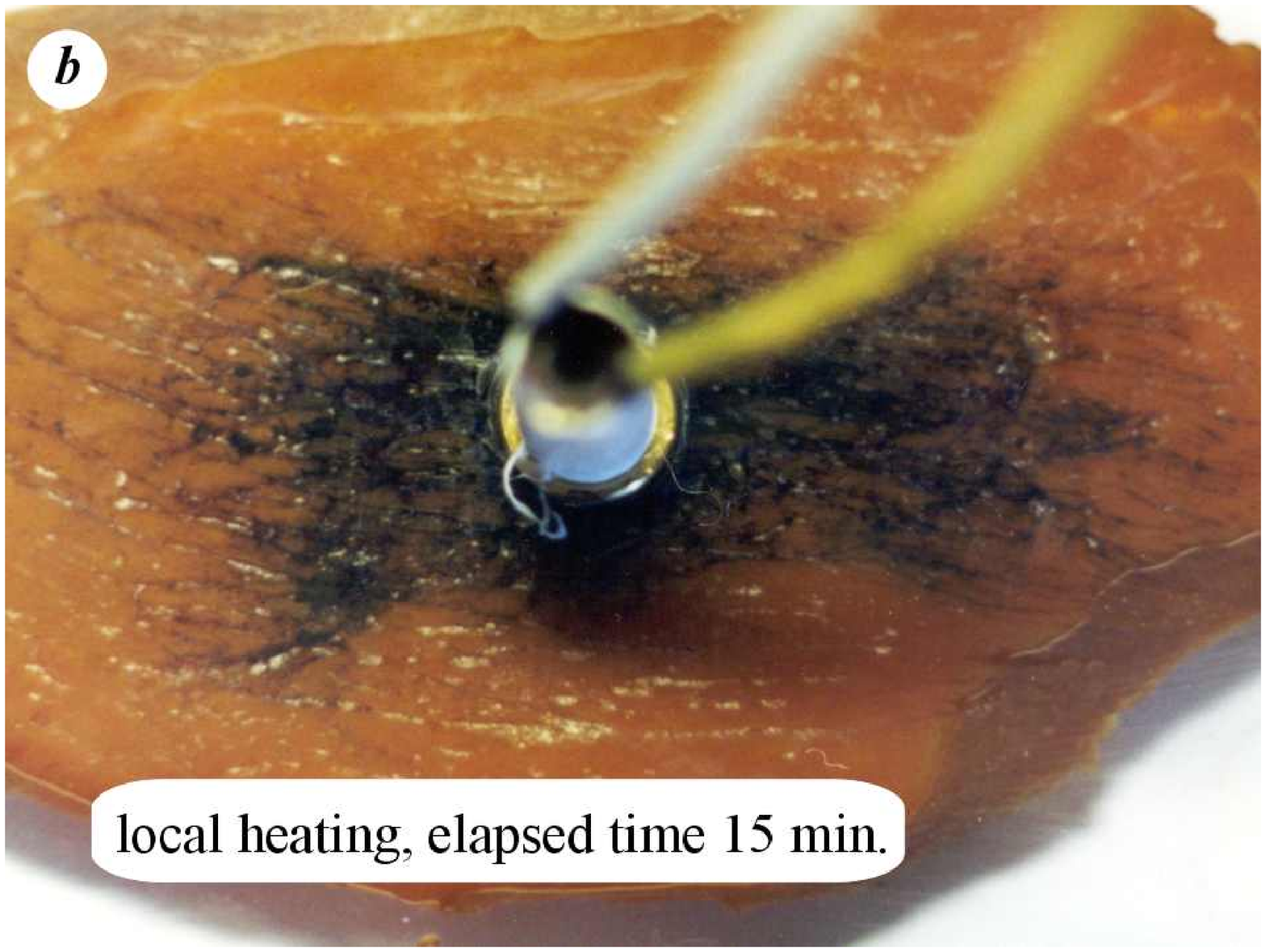}
\end{center}
\caption{Colour patterns formed by aniline dye spreading over the tissue from
the initial small drop (Fig.~\protect\ref{F6}) without heating ($a$) and under
strong local heating ($b$). (In the this figure the initial drop of aniline dye
located under the heating source is invisible.)\label{F7}}
\end{figure}

Figure~\ref{F6} illustrates the experiment showing the effect of nonuniform
heating on propagation of aniline dye in the cellular tissue. We put an aniline
dye drop in a lunula on the horizontal tissue surface and pressed down the
tissue with a glass plate to make the tissue surface plane and, thereby, to
exclude the influence of the surface relief. A metallic disc of diameter of
1\,cm was embedded into the glass plate placed so that this disc be right over
the drop. An ohmic resistance was attached to the external disc surface and
passing a current through it we locally heated the tissue. The colour patterns
formed by the aniline propagation are represented in Fig.~\ref{F7}.
Figure~$(a)$ exhibits the colour pattern appeared in the tissue without
heating after a lapse of 8 hours. As is seen, it is roughly of symmetrical
form and the muscle fibers have no substantial effect on the aniline
propagation in this case. The opposite situation is shown in figure~$(b)$ for
the aniline propagation under a strong local heating. Already after 15~min the
colour pattern attained the previous one in size and got a substantially
anisotropic form, the pattern dimension along the fibers is more that twice
its mean size in the transverse direction. Moreover, in this figure it is
clearly seen as aniline dry propagates inside small grooves on the tissue
surface whereas without heating such grooves practically have no effect.

\section{Discussion and hypotheses \protect\label{sec:discuss}}

The presented preliminary experimental data enable us to suggest that heat
transfer in cellular tissue under local strong heating is a more complex
phenomenon than a simple heat diffusion. On one hand, the observed strong
anisotropy in the heat propagation along the muscle fibers and in the
transverse direction cannot be explained keeping in mind only the heat
conduction through the cellular tissue. Indeed, otherwise, we should assume
that the temperature diffusivity along the fibers, $D_l$, is fully three times
greater than the corresponding value in the transverse direction, $D_t$
($D_l\gtrsim D_t$), whereas the cellular tissue contains no elements differing
from one another so remarkably in the heat characteristics. On the other hand,
the fiber arrangement endows the muscle tissue with a significant anisotropy
for motion of the interstitial liquid through the tissue as through a porous
medium. Therefore, if the nonuniform temperature field can cause the
interstitial liquid to flow then heat propagation in the muscle tissue will be
able to exhibit substantial anisotropy in properties. We think that in the
presented experiments exactly this phenomenon, i.e., the interstitial liquid
flux induced by nonuniform heating and, in turn, its effect on heat transfer
was observed. Obviously, it should be nonlinear, which corresponds to the data
presented in Fig.~\ref{F2}.

The experimental data shown in Fig.~\ref{F3} enable us to assume that heat
transfer in cellular tissue under such conditions is characterized by the
repeat reversibility. The latter means that the time variations in the
temperature of a cellular tissue undergoing repeated acts of heating go
approximately in the same way. So a possible microscopic mechanism responsible
for the induced flux of the interstitial liquid is also of the reversible
nature, at least, it is not due to the tissue thermal coagulation or another
thermal damage of the tissue. For example, it could be implemented through the
liquid redistribution between the tissue cells and the intercellular space due
to the change in the osmotic equilibrium induced by heating. In the given case
an additional amount of the interstitial liquid appearing in the heated region
has to give rise to its flow from this region.

We point out again that the interstitial liquid flux induced by heating seems
to be the integral feature of heat transfer in cellular tissue. Therefore, only
under special conditions heat propagation in cellular tissue is reduced solely
to diffusion. Moreover, mass and heat propagation have to be related with each
other through this flux, which is illustrated in Figs.~\ref{F5} and \ref{F7}.

Concluding the paper we would like to note that we observed anomalous
properties of heat transfer in muscle tissue where the fibers endow the tissue
with well developed anisotropy. There are organs, for example, kidney and
liver whose cellular tissue does not possess significant anisotropy on the
average regarding such an organ as a whole. For rigorously isotropic media the
effect of a similar liquid flux induced by heating is likely to be depressed.
However, on smaller scales these organs are certainly to exhibit local
anisotropy with random orientation in space. Therefore, local strong heating
again has to induce the interstitial liquid flux that, may be, changes its
direction randomly in different regions of a given organ. In this case heat
transfer is also to exhibit anomalous properties, which cannot be described, in
general, by the conventional diffusion equation (see, e.g., \cite{I92}).

\begin{acknowledgments}
These investigations were support in part by the Russian State Programme
``Integration'', Grant À0075.
\end{acknowledgments}

\end{document}